\documentclass[aps, prl, amssymb, amsmath, superscriptaddress, 
twocolumn] {revtex4-2}

\usepackage{chngcntr}

\usepackage{graphicx}
\usepackage{color}
\usepackage{amsmath}
\usepackage{enumitem}
\usepackage{amssymb}
\usepackage{hyperref}
\usepackage{cancel}
\usepackage{ulem}
\usepackage{multirow}

\newcommand{\be}{\begin{equation}}
	\newcommand{\ee}{\end{equation}}

\newcommand{\bea}{\begin{eqnarray}}
	\newcommand{\eea}{\end{eqnarray}}

\newcommand{\p}{\partial}

\newcommand{\la}{\left\langle}
\newcommand{\ra}{\right\rangle}
\newcommand{\lb}{\left[}
\newcommand{\rb}{\right]}
\newcommand{\lp}{\left(}
\newcommand{\rp}{\right)}

\renewcommand{\vec}[1]{{\boldsymbol #1}}

\makeatother

\begin{document}
	\title{Collective excitations in chiral Stoner magnets}
	\author{Zhiyu Dong, Olumakinde Ogunnaike, and Leonid Levitov}
	\affiliation{Department of Physics, Massachusetts Institute of Technology, Cambridge, MA 02139}
	
	\begin{abstract}
		We argue that spin and valley-polarized metallic phases recently observed in graphene bilayers and trilayers support chiral edge modes that allow spin waves to propagate ballistically along system boundaries without backscattering. The chiral edge behavior 
		originates from the interplay between the momentum-space Berry curvature in Dirac bands and the geometric phase of a spin texture in position space.
		The edge modes are weakly confined to the edge, featuring dispersion which is robust and insensitive to the detailed profile of magnetization at the edge. This unique character of edge modes reduces their overlap with edge disorder and enhances the mode lifetime. The mode propagation direction reverses upon reversing valley polarization, an effect that provides a clear testable signature of 
geometric interactions in isospin-polarized Dirac bands.
	\end{abstract}
	
	\maketitle
	
\begin{figure}
		\centering \includegraphics[width=0.99\columnwidth]{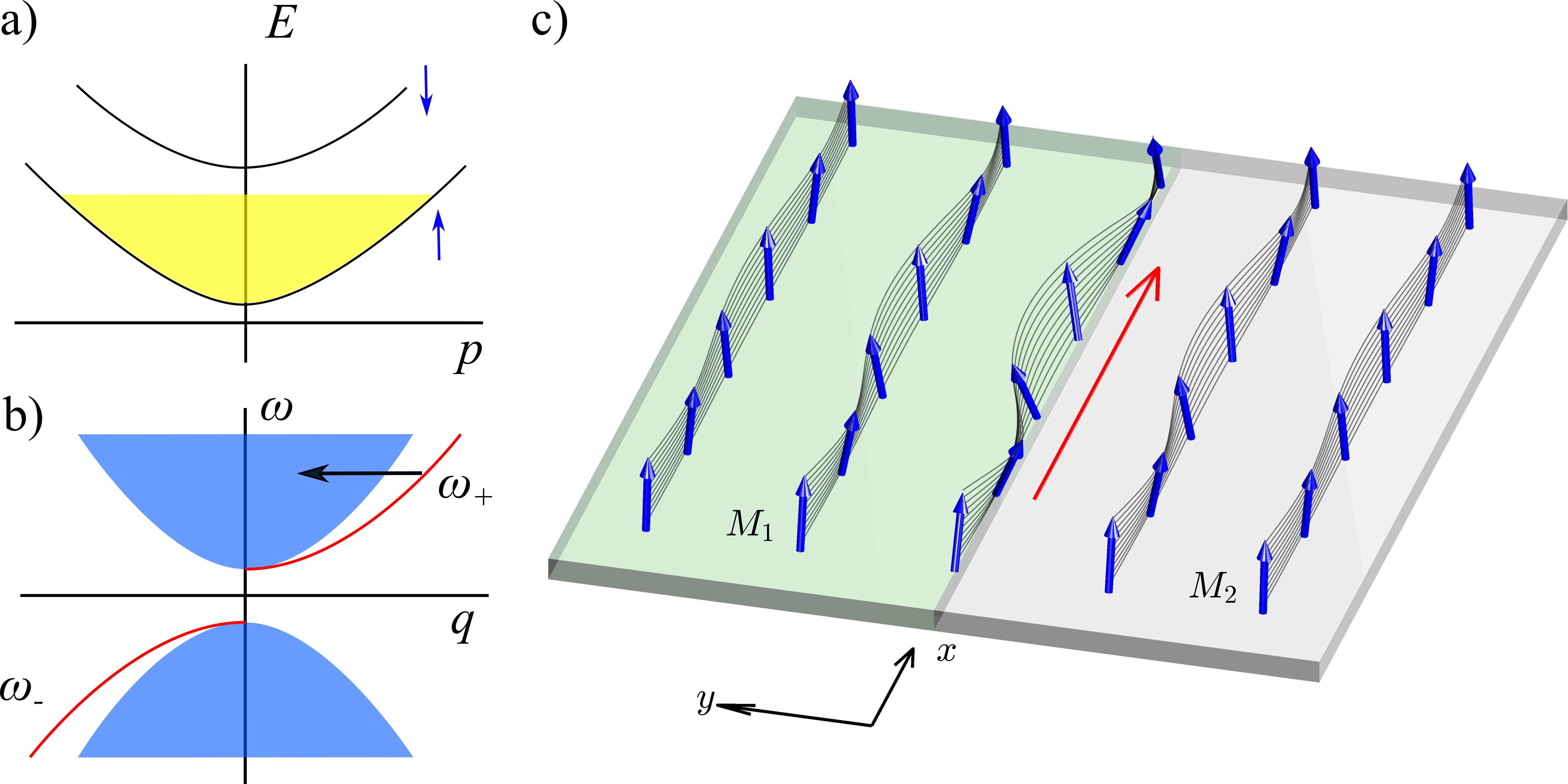}
		\caption{(a) Schematic band structure of a fully spin-polarized Stoner phase in a valley-polarized graphene bilayer or trilayer band. Only the valley populated by carriers is shown. 
			(b) The spin-wave edge mode dispersion obtained for a step in orbital magnetization $M_1\ne M_2$ induced by a gate, Eq.\eqref{eq:step_in_M}. The mode (red) is positioned outside the bulk magnon continuum (blue). The group velocity $v_g=d\omega/dq$ of a constant sign indicates the chiral character of the mode. The edge-to-bulk scattering (black arrow) is blocked by the energy and momentum conservation. (c) Schematic of the spatial dependence of the edge mode. The chiral mode is confined to the step and propagates along it without backscattering. 
		}
		\label{fig:edge_mode}
\end{figure}

Stoner ferromagnetism 
is a correlated electron order ubiquitous 
in topological materials of current interest, including moir\'e graphene\cite{andrei2020graphene,cao2018insulator,cao2018SC,zondiner2020cascade,wong2020cascade,saito2021isospin}, and nontwisted graphene bilayers  and trilayers 
\cite{zhou2022isospin,seiler2021quantum,de2021cascade,zhou2021half,
zhou2021superconductivity}. Yet, the fundamental properties of this state, especially those governed by Berry curvature in $k$ space,  
are presently poorly understood. Here we predict that this state hosts 
chiral spin excitations. These excitations are confined to system edges and domain boundaries between different valley-polarized regions, propagating along them in a manner resembling Quantum Hall (QH) edge states, as illustrated in Fig.\ref{fig:edge_mode}. The microscopic origin of this behavior is the geometric phase of carrier spins tracking magnetization along carrier trajectories.  
Carrier spin rotation by a position-dependent magnetization generates a Berry phase in direct space that serves as a spin-dependent magnetic vector potential 
that couples to the orbital dynamics of carriers (see Eqs.\eqref{eq:a},\eqref{eq:B(r)})  \cite{ohgushi2000spin,fujita2011gauge,nagaosa2012emergent,hamamoto2015quantized}. 
The chiral edge behavior arises due to the coupling between this geometric magnetic field and  orbital magnetization due to Berry curvature in $k$ space. The geometric character of this interaction ensures robust chiral edge physics even in ``vanilla'' spin-polarized Fermi seas such as those seen in Refs.\cite{zhou2022isospin,seiler2021quantum,de2021cascade,zhou2021half,zhou2021superconductivity}. 
		
The band magnetism of carriers exhibiting orbital magnetization is a broad framework applicable to a diverse range of systems. 
		This includes, in particular, the 
		QH ferromagnets 	\cite{girvin2000spin,nomura2006quantum,alicea2006graphene,yang2006collective} and correlated excitonic phases in 
		QH bilayers \cite{spielman2000resonantly,eisenstein2014exciton,eisenstein2004bose,
			li2017excitonic,finck2010quantum}. 
		Orbital magnetization in these systems exists due to 
		Landau levels 
		rather than the $k$-space Berry curvature and in QH bilayers the layer index plays the role of spin in our analysis. 
		Here we 
		focus on 
		chiral edges in spin-polarized metals and, afterwards, comment on possible extensions to the QH systems. 

	In graphene multilayers \cite{zhou2022isospin,seiler2021quantum,de2021cascade,zhou2021half,zhou2021superconductivity},  the predicted chiral edge behavior is 
	sensitive to valley polarization. 
	In a valley- and spin-polarized phase (identified as a quarter metal in Refs.\cite{zhou2022isospin,seiler2021quantum,de2021cascade,zhou2021half,zhou2021superconductivity}), the band orbital magnetization exhibits opposite signs in valleys $K$ and $K'$. As a result, the chirality (i.e., the propagation direction) of edge modes flips upon reversing the valley imbalance.
	
	A very different behavior is expected in a valley-unpolarized but spin-polarized phase (half-metal in the nomenclature of  Refs.\cite{zhou2022isospin,seiler2021quantum,de2021cascade,zhou2021half,zhou2021superconductivity}). In this case, the two valleys host Stoner metals with the band orbital magnetization of opposite signs. In this phase, the edges will host pairs of counter-propagating chiral edge modes, one for each valley.  These two modes together respect the orbital time reversal symmetry, unbroken in the half-metal phase, i.e. the system is non-chiral. 

The exceptional cleanness of graphene multilayers makes them an appealing system to probe this behavior. Spin lifetimes as long as 6 ns measurered in large bilayer graphene (BLG) systems by a nonlocal Hanle effect at 20 K \cite{han2012spin} are explained by residual magnetic disorder \cite{kochan2014spin,kochan2015resonant}. In contrast, recently, it was demonstrated that electrons isolated from edge disorder by gate confinement and trapped in gate-defined quantum dots acquire ultralong spin lifetimes, reaching values of 200 ${\rm \mu s}$ \cite{banszerus2022spin} and $50$ ms \cite{gachter2022single} when measured in an applied magnetic field by pulsed-gate spectroscopy. Therefore, probing spin excitations in gate-defined electron puddles presents a distinct advantage. Yet, spin lifetimes measured in large BLG systems \cite{han2012spin} also lie in a suitable range. Spin lifetimes can be further increased by applying nonquantizing magnetic fields that, apart from a constant offset, have little impact on the chiral spin-wave dispersion (see Eq.\eqref{eq:dispersion}).

In a metallic state the chiral mode at the edge can, in principle, decay 
by scattering into the 2D spin-one particle-hole continuum and spin waves. 
The former process is blocked by energy conservation since the spin-one continuum is gapped at small momenta [see 
Fig.\,\ref{fig:edge_mode_alt} (a)].
The latter process, 
as shown by the black arrow in Fig.\ref{fig:edge_mode} (b), is 
	blocked by the energy and momentum conservation
for a smooth edge but can be viable for a rough edge. However, as discussed in \cite{SM}, in the long-wavelength limit the edge modes have vanishing overlaps with the edge disorder potential, a property that protects the modes from edge-to-bulk scattering. 



The chiral edge behavior in a Stoner metal phase discussed here is distinct from that predicted 
for magnetic phases with a nontrivial magnon band topology 
\cite{mook2014edge, maeland2022quantum,diaz2019topological,garcia2014nonreciprocal,
mcclarty2022topological,shindou2013topological,shindou2013chiral}. 
In these systems, chiral edge excitations lie above the first magnon band 
and are therefore gapped. To the contrary, the chiral modes described here 
arise at the boundary of a uniformly spin-polarized Stoner Fermi sea---a metallic compressible state with a nontopological bulk magnon band. 
The edge excitations are gapless (in the absence of an externally applied magnetic field, see below) and have dispersion positioned beneath that of bulk spin waves (in our case these are nothing but the gapless magnons of a Heisenberg ferromagnet). 
Accordingly, here chiral modes arise in the absence of microscopic spin-dependent interactions such as Dzyaloshinskii-Moriya interaction (DMI) or dipolar interaction (as in Refs.\cite{mook2014edge,maeland2022quantum,diaz2019topological,
garcia2014nonreciprocal,mcclarty2022topological} and Refs.\cite{shindou2013topological,shindou2013chiral}, respectively). Instead, they originate from an interplay between the exchange interaction and orbital magnetization in bands with Berry curvature and broken time reversal symmetry. Our spin waves 
act analogously to the chiral edge plasmons predicted for such bands \cite{song2016chiral}, yet they transport spin rather than charge and arise from a very different mechanism.



Collective spin dynamics, both bulk and edge, are readily analyzed in the long-wavelength limit, at frequencies 
below the Stoner continuum [see 
Fig. \ref{fig:edge_mode_alt} (a)]:
\be\label{eq:Stoner_gap}
\Delta =U n_s>\omega(q)
,
\ee 
where $\Delta$ is the Stoner gap, $U$ is the exchange interaction, $n_s$ is spin-polarized carrier density and $\omega(q)$ is mode dispersion.
We employ an effective action for spin variables obtained by integrating out fermion orbital degrees of freedom. In that, we assume the electron velocity is large compared to that of spin-waves, $v_F\gg v_g=d\omega/dk$. As found below, the long-wavelength spin-wave dispersion is quadratic, $\omega(k)\sim k^2$, a behavior that confirms the separation of time scales for the orbital and spin degrees of freedom and justifies our analysis. 
The effective action for spin variables takes the form  [see e.g. \cite{nagaosa1999quantum,fradkin2013field}]
\begin{align}
	A= \int dt d^2 r \lp i n_sS_0\langle\eta(\vec r,t)|
	\p_t|\eta(\vec r,t) \rangle- \mathcal  H[\vec n]\rp
	,
	\label{eq:action}
\end{align}
where the first term is the Wess-Zumino-Witten action, hereafter referred to as $A_{\rm WZW}$, representing the single-spin Berry phase accumulated through time evolution. The second term is the Hamiltonian of a spin-polarized state discussed below. 
The quantity $|\eta(\vec r,t)\rangle$ represents a coherent spin state in (2+1)D space-time.
Here $n_s=n_\uparrow-n_\downarrow$ is the density of spin-imbalanced carriers, 
the factor $n_s S_0$ is the spin density, where $S_0=\hbar/2$. In what follows spin polarization is described by a unit vector 
\[
\vec n(\vec r,t) = \langle \eta(\vec r,t)|\vec \sigma| \eta(\vec r,t)\rangle.
\] 
The term $\mathcal H[\vec n]$ in Eq.\eqref{eq:action} is the effective spin  Hamiltonian. 
Symmetry arguments and microscopic analysis predict \cite{dong2022chiral} the long-wavelength Hamiltonian 
\be\label{eq:H}
\mathcal H[\vec n] =  n_s \lb \frac{J}{2}(\partial_\mu \vec n)^2 - M(r) B(\vec r,t)
- \vec h_0\cdot\vec n \rb
.
\ee
Here $J$ is spin stiffness, the second term
is an interaction between the band orbital magnetization and the geometric magnetic field, the last term 
is the Zeeman energy per carrier, with  the $g$-factor and Bohr magneton absorbed in 
the external magnetic field $\vec h_0$.

As indicated above, the interaction $-MB$ originates from a geometric Berry phase, arising 
	due to electron spins tracking magnetization along electron trajectories. Spin rotation generates a Berry phase in position space defined by a spin-dependent magnetic vector potential  \cite{ohgushi2000spin} 
	\be\label{eq:a}
	a_\mu = \frac{\hbar c}{2e}\lp1-\cos\theta\rp \partial_\mu \phi,\quad \mu=x,y
	.
	\ee
	Here $\theta$ and $\phi$ are the polar and azimuthal angles measured with respect to the spin polarization axis in the ground state. The sign of $a_\mu$ is chosen to describe the Berry phase accrued by the majority-spin carriers. For the minority-spin carriers the vector potential is of the opposite sign and is described by $-a_\mu$, giving a Berry phase of the opposite sign.
	The geometric magnetic field is simply the curl of $a_\mu$. In terms of $\vec n$, it reads: 
	\be\label{eq:B(r)}
	B(\vec r,t)=\nabla\times \vec a=\frac{\phi_0}{4\pi}\vec n\cdot (\partial_{x}\vec n\times \partial_{y}\vec n),
	\ee 
	where $\phi_0=hc/e$ is the flux quantum.
	This physics was first discussed 
	in the early literature on high $T_c$  superconductivity\cite{baskaran1988gauge,wiegmann1988superconductivity,schulz1990effective,ioffe1991hall} and later in the literature on noncollinear magnetic systems \cite{ohgushi2000spin,fujita2011gauge,hamamoto2015quantized,nagaosa2012emergent}. Importantly, unlike static spin textures in the latter systems, our spin-wave dynamics generate a time-dependent vector potential, Eq.\eqref{eq:a}. This yields a geometric electric field \cite{nagaosa2012emergent,back20202020} 
	\be\label{eq:E(r)}
	E_\mu=-\partial a_\mu /c \partial t-\nabla a_0
	=\frac{\hbar}{2 e}\vec n\cdot (\partial_{t}\vec n\times \partial_{\mu}\vec n),
	\ee
	which 
	can enable electrical detection of the spin waves.

The quantity $M(r)$ in the 
	second term in Eq.\eqref{eq:H} 
	describes the orbital magnetization 
	per carrier in a spin-imbalanced band arising  due to Berry curvature in $k$ space. 
It is given by a sum of contributions of the filled states in the spin-valley-polarized Fermi sea. For a partially spin-polarized Fermi sea the contributions to $M$ 
from the majority-spin and minority-spin carriers are of opposite signs, 
	giving $M=M_{\uparrow}-M_{\downarrow}$. 
	The opposite signs originate from the opposite signs 
	of $a_\mu$ for the spin-up and spin-down carriers discussed beneath Eq.\eqref{eq:a}. 
These opposite sign contributions cancel in a spin-unpolarized state but lead to $M\ne 0$ in a fully or partially spin-polarized state. 
The position dependence $M(r)$ reflects spatially varying spin or valley imbalance arising, e.g., due to gating. 

The geometric fields $a_\mu$, $B$ and $\vec E_\mu$ are  derived in the adiabatic regime when an electron spin tracks spin texture along the electron's trajectory. 
The adiabatic regime occurs when the spin texture is sufficiently long-wavelength such that the Stoner spin gap $\Delta=Un_s$ is much greater than $\hbar v_F q$, where $q$ is the characteristic spin-wave wavenumber and $U$ is the exchange interaction (see Eq.\eqref{eq:Stoner_gap}).

	
	The Hamiltonian, Eq.\eqref{eq:H}, features different 
	phases 
	depending on the $M$ and $J$ values \cite{dong2022chiral}. If $M>2J$ and $h_0$ is small enough, the uniformly polarized state is predicted to become unstable towards twisting, giving rise to a skyrmion texture with a nonzero chiral density $B$. Here, we consider 
	excitations in 
	a uniformly polarized state 
	\be\label{eq:linearization S=S+dS}
	\vec n(r,t)= \vec n_0+\delta\vec n(r,t),\quad
	 \delta\vec n\perp \vec n_0
	 ,
	\ee 
	with $\vec n_0\parallel \vec h_0$,  
	occurring for not too large $M$ values. 

	The spin wave dispersion can be obtained from the canonical equations of motion 
	found from the saddle-point condition $\delta A/\delta \vec n = 0$, with $A$ given in Eq.\eqref{eq:action}. 
	Indeed, the variation of the Wess-Zumino-Witten term $A_{\rm WZW}$ [the first term in Eq.\eqref{eq:action}] can be found by noting that this term equals to $n_s S_0$ times the solid angle swept by $\vec n$. As a result, its variation can be expressed as
	\be
	\delta A_{\rm WZW} = n_s S_0 \int dt d^2r (\delta \vec n \times \p_t\vec n) \cdot \vec n,
	\ee
	The variation of the action in Eq.\eqref{eq:action} gives
	$\delta A = \lp n_s S_0 \p_t \vec n\times \vec n-\delta {\cal H}/\delta \vec n\rp\cdot \delta \vec n$,
	giving equations of motion:
	\be
	n_s S_0 \partial_t \vec n(r) = \vec h(r)\times \vec n(r),\quad \vec h 
	= -\frac{\partial \mathcal H}{\partial \vec n} + \partial_\mu \frac{\partial \mathcal H}{\partial \partial_\mu \vec n}.
	\ee
	Linearizing about a uniformly polarized state 
	yields coupled linear equations for $\delta \vec n$ components, which are identical to those found for a nonchiral problem,
	\be S_0 \partial_t \delta \vec n(r,t) = \vec h_0\times \delta \vec n(r,t)+ J\partial_\mu^2 \delta \vec n(r,t)\times\vec n_0. 
	\ee
	Plane wave solutions to this equation yield a simple isotropic and non-chiral spin-wave dispersion
	\be\label{eq:spinwaves2D}
	\omega_\pm(q) = \pm (h_0+Jq^2)/S_0,
	\ee
	with 
	values approaching $\pm h_0/S_0$ in the limit $q\to 0$, universally and independent of the exchange interaction, as required by the Larmor theorem.
	
%
	%
	
For a spatially uniform $M$, the $-MB$ term is a topological invariant. Therefore, a local twist of spin does not change the ${\cal H}$ value. As a result, this interaction neither affects the energy nor impacts the spin waves. A spatially varying $M$, to the contrary, has a profound effect on spin waves. In particular, system boundaries and interfaces between regions in which $M$ takes different values support chiral spin-wave modes 
reminiscent of the QH edge states. 
	To illustrate this behavior we consider a step
	\be\label{eq:step_in_M}	
	M(y) = \begin{cases}
		M_1,\quad y>0\\
		M_2,\quad y<0.
	\end{cases}
	\ee
	In this case, after linearization, Eq.\eqref{eq:linearization S=S+dS}, we find
	\be
	\vec  h = n_s \lb  J\partial_\mu^2 \delta \vec n 
	- \partial_y M(y) (\vec n_0 \times \partial_x \delta \vec n)
	+\vec h_0 \rb
	.
	\ee
	Other terms vanish at first order in $\delta \vec n$. 
	As a result, the linearized equations of motion become 
	\be
	S_0\partial_t \delta \vec n = \vec h_0\times \delta \vec n+J  \partial_\mu^2 \delta \vec n\times \vec n_0 + m\delta(y)   (\vec n_0 \times \partial_x \delta \vec n)\times\vec n_0,\nonumber
	\ee
	where $m=M_2-M_1$ is the difference between $M$ on two sides of the edge. 
		These equations are solved by writing $\delta \vec n(x,y)$ as a superposition of complex-valued 
		helical components:
		\begin{align}\nonumber
			\delta \vec n(r,t) =\lp\begin{matrix}
				\delta n_x(r,t)\\
				\delta n_y(r,t)
			\end{matrix}\rp 
			= & \sum_{q} e^{iqx} 
			\lb e^{-i\omega_{+}t} \psi_{q,+}(y)
			\lp\begin{matrix}
				1\\
				i
			\end{matrix}\rp \right.
			\\
			& \left. +
			e^{-i\omega_{-}t} \psi_{q,-}(y)
			\lp\begin{matrix}
				1\\
				-i
			\end{matrix}\rp\rb ,
		\end{align} 
		where 
		we
		carried out the Fourier transform in time and the translation-invariant $x$ direction. 
		Plugging this ansatz into the equations of motion for $\delta \vec n(r,t)$, we obtain two decoupled 1D problems for a quantum particle in a delta-function potential, separately for each helicity:
		\be
		S_0\omega_{\pm} \psi (y) = \pm
		\lb h_0 +J (q^2-\partial_y^2) \rb  \psi -
		mq\delta(y) \psi (y),
		\ee
where $\psi(y)$ is a shorthand for $\psi_{q,\pm}(y)$.
		These equations support bound states which are edge spin waves for the helical polarization of a plus (minus)  sign 
		for $mq$ of a positive (negative) sign, respectively. 
		
		Indeed, the bound state is described by an exponential solution for both helicities: 
		\be\label{eq:psi_pm}
		\psi_{q,\pm}(y) = u_q e^{-\lambda_q |y|}, \quad \lambda_q >0,
		\ee
		where the condition $\lambda_q >0$ is required for the mode to be normalizable.
		The value of $\lambda_q$ and the dispersion are determined by the condition 
		\be\label{eq:omega+-}
		0 = \pm
		2J \lambda_q \delta(y)  - 
		mq\delta(y),
		\ee
which gives $\lambda_q = \pm\frac{mq}{2J}$. Therefore, 
the right-helicity mode $\psi_{+}^q$ exists only for $mq>0$, whereas the left-helicity mode $\psi_{-}^q$ exists only for $mq<0$. 
		\be\label{eq:dispersion}
		\omega_\pm(q) 
		=  
		\pm\frac{1}{S_0}\lb h_0+ \lp J-\frac{m^2}{4J}\rp q^2\rb 
		\ee
		The resulting dispersion is illustrated in Fig.\ref{fig:edge_mode} (b) for $m>0$. The group velocity $v_g=d\omega/dq$ is of the same sign for both helicities, as expected for a chiral edge mode. At $q=0$, the frequency value agrees with the Zeeman frequency for a single spin, as required by Larmor's theorem. At this point $\lambda_q$ vanishes, which signals that the mode ceases to be confined to the edge and transforms into a uniformly precessing state.
		
		Notably, the discrete chiral mode, Eq.\eqref{eq:dispersion}, appears in a robust manner regardless of magnetization values 
		in the two halfplanes and the step size $m=M_1-M_2$. At $M_1$ approaching $M_2$ the chiral mode, while remaining discrete, approaches the bulk magnon continuum and merges with it at $M_1=M_2$. Another interesting aspect of the 
		dispersion in Eq.\eqref{eq:dispersion} is that the group velocity reverses when $m$ exceeds $2J$, upon which the mode 
		propagation direction is reversed, with the left-moving excitations becoming right-moving and vice versa. In this regime the frequencies $\omega_\pm(q)$ reverse their signs when the wavenumber reaches a certain critical value, $q=q_*=\sqrt{4Jh_0/(4J^2-m^2)}$. Frequency sign reversal signals an instability towards a spatial modulation at the edge 
		with spatial periodicity $2\pi/q_*$. 
		Notably, this instability can occur before skyrmions are nucleated in the bulk. This happens, in particular, when $M_1$ and $M_2$ are of opposite signs. In this case, the condition for skyrmion nucleation in the bulk, $2J<|M_{1,2}|$, is more stringent than that for the instability at the edge, $2J<|M_1-M_2|$.

Next, we consider polarization of chiral modes. As we found above, the modes of both helicities, $\psi_+$ and $\psi_-$, propagate in the same direction. This gives rise to an interesting space-time picture that combines propagation with velocity $v_g$ and precession about $\vec h_0$. Indeed, a narrow wavepacket $u_q$ centered at $q\approx q_0$ evolves as
		\begin{align}\label{eq:delta S}
			\delta \vec n(r,t)&=\sum_{q>0}  
			\phi^+_{q}(r,t) \lp\begin{matrix}
				1\\
				i
			\end{matrix}\rp 
			+ \sum_{q<0}\phi^-_{q}(r,t) \lp\begin{matrix}
				1\\
				-i
			\end{matrix}\rp 
				\\ \nonumber
				&\sim e^{-\lambda_{q_0} |y|}
				u(x-v_gt)
				\lp\begin{matrix}
					\cos\lb \omega_0t-q_0x 
					+\theta_0\rb\\
					\sin\lb \omega_0t-q_0x 
					+\theta_0\rb 
				\end{matrix}\rp 
				.
			\end{align}
			Here, $\phi^\pm_{q}(r,t)=e^{-i\omega_\pm (q)t+iqx-\lambda_q|y|}u_q$. The quantity $u(x)$ is the Fourier transform of $u_q$, 
			$\omega_0=\omega_+(q_0)$, $v_g$ is the group velocity $d\omega/dq$ at $q=q_0$, $\theta_0$ is a free parameter.
			This describes 
			spin precession and 1D propagation, 
			as illustrated in Fig. \ref{fig:edge_mode} (c). 
			
Lastly, we discuss the relation between the analysis above and 
				the collective spin excitations in QH 
				ferromagnets. The seminal prediction of skyrmions in QH ferromagnets by Sondhi et al.\cite{sondhi1993skyrmions} relies on the notion of an excess charge induced on a chiral spin texture, $\delta \rho(r) = \frac1{c}\sigma_{xy} B(r)$, a value that follows from the topological pumping argument \cite{thouless1983quantization,niu1984quantised} with $\sigma_{xy}$ the Hall conductivity of a filled Landau level and $B$ the quantity in Eq.\eqref{eq:B(r)}. 
				This gives a contribution to the energy 
				\be\label{eq:charging_Landau_level}
				\delta E = \int d^2r  V_g \delta \rho(r),
				\ee
				where $V_g$ is the gate voltage. 
				Since $B(r)=\frac{\phi_0}{4\pi}\vec n\cdot \p_1 \vec n\times \p_2\vec n$, the quantity in Eq.\eqref{eq:charging_Landau_level} is identical in form to our $-MB$ interaction (the second term in Eq.\eqref{eq:H}). Furthermore, it is straightforward to link the prefactor with the orbital magnetization of a fully filled Landau level 
				\be 
				M = \frac1{c}V_g\sigma_{xy}.
				\ee
This relation follows from the thermodynamic relation $ dM/d\mu =dn/dB_{\rm ext}$ and the Streda formula $dn/dB_{\rm ext} =  \frac{\sigma_{xy}}{ce}$. Having reproduced the $-MB$ interaction in the QH framework, 
we are led to conclude that the chiral spin waves derived above must also occur in QH ferromagnets.
While a detailed analysis should be deferred to future work, we expect that these modes differ in two distinct ways from various chiral charge and spin edge modes that have been widely investigated in QH systems 
\cite{balaban1997observation,mazo2012collective,tikhonov2016emergence,iordanski2002excitations,karlhede1999dynamics,zhang2013edge,kharitonov2016interplay,saha2021emergence,khanna2022emergence}First, their dispersion at small $k$ will be quadratic rather than linear. Second, rather than being tightly confined to the edge on a magnetic length scale, these modes will feature a wider profile extending far into the bulk. The weak confinement may suppress scattering by edge disorder and boost the lifetimes for these modes.

Last, we envision that extending the pulsed gate spectroscopy of Refs. \cite{banszerus2022spin,gachter2022single} to probe the gate-confined electron puddles can allow to launch the chiral spin waves and detect them in a manner analogous to the time-domain detection of QH edge magnetoplasmons \cite{ashoori1992edge,zhitenev1993time,ernst1996acoustic,kumada2011edge}. Further, 
electron-spin resonance (ESR) measurements on such puddles by the technique recently used to probe ESR in graphene\cite{sichau2019resonance} can provide direct information of the chiral mode dispersion.  Indeed, for a puddle of circumference $L$ the mode dispersion 
in Eq.\eqref{eq:dispersion}, will translate into sidebands of the ESR resonance with frequencies
\be
\omega_n=\omega(q_n),\quad q_n=2\pi n/L,
\ee with integer $n$. 
Here $n=0$ is the fundamental ESR frequency and
$n=1,2,3...$ describes a family of chiral mode excitations. The $\omega=\omega_n$ resonances will occur over a continuous background due to the 2D spin-wave continuum, Eq.\eqref{eq:spinwaves2D}. 
As an example, we consider a disk of circumference $L=10\,{\rm \mu m}$ for which the minimal wavenumber is $q_1=2\pi/L$. Estimating the stiffness as the e-e interaction at the Fermi wavelength scale, $J\sim e^2/(\kappa \lambda_F)$, and plugging realistic parameter values, we find the sideband frequency detuning of $\omega_1-\omega_0\approx 50$ MHz. This value is greater than $1/T_1$ found in Refs.\cite{banszerus2022spin,gachter2022single} and lies in a convenient spectral range for microwave measurements. We also note that, as discussed above, spin dynamics in our system is accompanied by a geometric electric field given in Eq.\eqref{eq:E(r)}. The oscillating electric polarization induced by this field can be used for a direct electrical detection of the chiral spin-wave dynamics. 

Summing up, the chiral edge excitations are a unique manifestation of geometric interactions  
in a metallic spin-polarized Fermi sea with a Berry band curvature.
Despite occurring in a non-topological setting they 
are protected from backscattering by their chiral character. Correlated-electron phases that host chiral edge modes allowing excitations to propagate along system boundaries in a one-way manner are of keen interest for fundamental physics and are expected to harbor interesting applications. We describe the requirements for such modes to exist and argue that the chiral behavior and associated exotic physics are generic and readily accessible in state-of-the-art systems.

This work originated from fruitful discussions with Eli Zeldov. We thank Herbert Fertig, Steven Girvin, Bertrand Halperin, Efrat Shimshoni, Shivaji Sondhi and Kun Yang for useful comments on the preliminary version of this paper. This research was supported by the Science and Technology Center for Integrated Quantum Materials, National Science Foundation Grant No. DMR1231319.

			\bibliography{ref_chiral_edge_mode}

\newpage

\section{Supplementary information} 
			
			Here, we consider intrinsic mechanisms of the edge spin wave damping. We first discuss 
			the Landau damping due to the 2D particle-hole continuum
ignoring edge roughness, and then consider the decay pathway that is enabled by the edge roughness. 
			We argue that both mechanisms give damping that becomes negligible at long wavelengths and low frequencies. 
			
			The particle-hole continuum in spin-polarized metals, that can potentially lead to Landau damping of collective spin excitations, consists of two distinct parts: 
			the spin-zero continuum and the spin-one continuum. The spin-zero particle-hole continuum spans wavenumbers $0<k<2k_F$ and extends in frequency down to $\omega =0$. Spin waves cannot simply decay into these excitations owing to the spin U(1) symmetry that ensures spin conservation. Therefore, these excitations do not impact the spin-wave lifetimes on a tree level. Scattering involving spin-zero continuum can only take place through higher-order processes in which a spin wave is scattered by a spin-zero excitation or emits it without decaying. However, such processes are suppressed by the phase space volume for the final states. 
			
\begin{figure}
\centering \includegraphics[width=0.99\columnwidth]{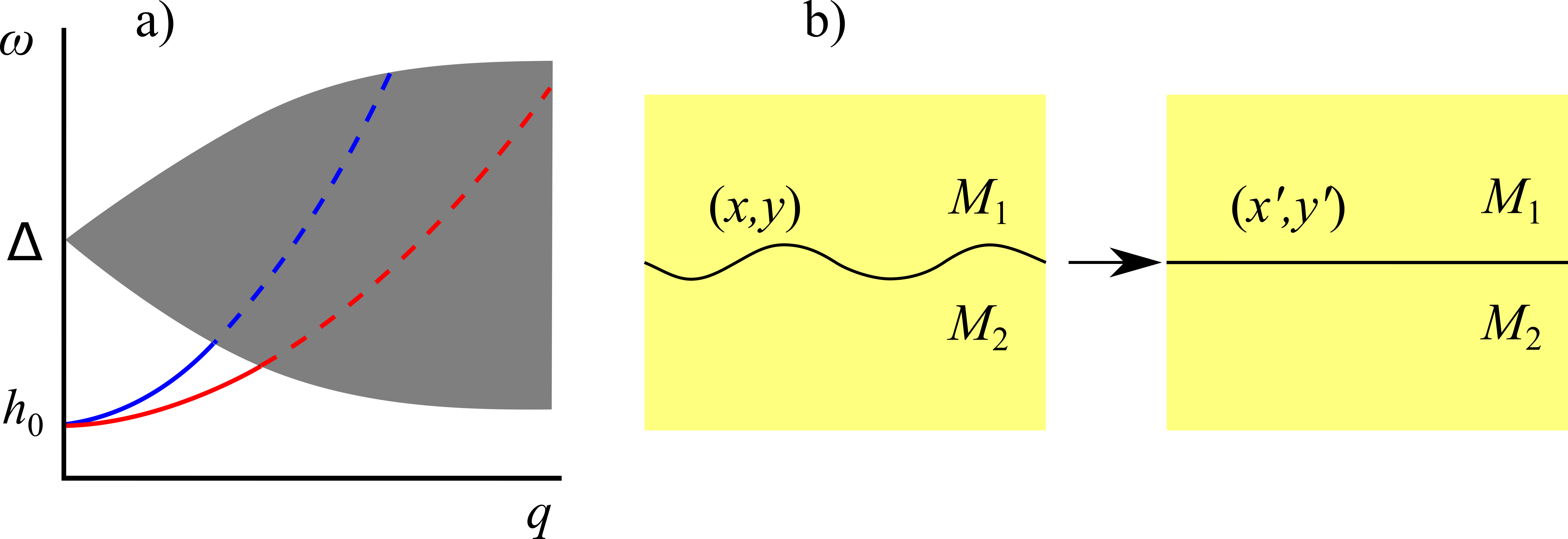}
\caption{(a) Collective spin excitations in the bulk (blue line) and at the edge (red line) superimposed with the spin-one particle-hole continuum (gray region). Below the Stoner gap $\Delta=Un_s$ the collective modes are decoupled from the continuum and are therefore discrete. Upon entering the continuum they become Landau-damped (dashed lines). The spin-zero particle-hole continuum with wavenumbers 
$0<k<2k_F$ and frequencies extending down to $\omega =0$ is not shown. This continuum is irrelevant for the damping of spin waves since 
one-excitation processes are blocked by U(1) spin conservation. (b) A wiggly step in magnetization $M(r)$ representing a rough edge. By a conformal mapping, Eq.\eqref{eq:g_k}, the problem with a wiggly edge is mapped onto the one with a straight edge and a fluctuating metric localized near the edge, Eq.\eqref{eq:metric}. The 
edge-to-bulk scattering gives rise to a finite lifetime of the edge mode, $\tau=1/\gamma$, 
where $\gamma$ is given in Eq.\eqref{eq:gamma}.
}
\label{fig:edge_mode_alt}
\end{figure}

			Another 2D particle-hole continuum nominally available for decay, which is not blocked by the spin conservation, is the spin-one continuum, in which an electron is excited from the spin-majority band to the spin-minority band. However, the spin-one excitations are fully gapped at small momenta [see Fig.\ref{fig:edge_mode_alt} (a)] and, as a result, this scattering pathway is absent for long-wavelength (low-frequency) spin waves. Moreover, for the fully spin-polarized phase, the spin-one excitation is fully gapped at all momenta, which fully protects the long-wavelength spin waves from the edge-to-bulk scattering.
						
			Next, we study the scattering from the spin wave edge mode to bulk spin waves by edge disorder, and show that the lifetime is ultra-long for long-wavelength edge modes. As a simple model, we consider a step in magnetization with a wiggly boundary, illustrated in Fig. \ref{fig:edge_mode_alt} (b). We will describe spin waves by a problem linearized in a weak perturbation about the uniform state, as implemented in the main text [see Eq.(7) therein]. Using the right and left helicity representation [see Eq.(14) of the main text] and, without loss of generality, focusing on the $\psi_+$ mode, we arrive at an effective action
			\begin{align}\label{eq:action_wiggly}
				A=\int dt d^2r 
				\bar\psi_+\lp i\p_t-H
				\rp \psi_+
				,
			\end{align}
			where $H=h_0-\frac{J}2(\p_x^2+\p_y^2)+\epsilon_{jj'}\p_j M(r)\p_{j'}$, 
			and $M(r)$ describes two domains with magnetization $M_1$ and $M_2$ [Fig.\ref{fig:edge_mode_alt} (b)]. The magnetization gradient $\p_j M(r)$ is a delta function centered at the wiggly domain boundary. 
			For conciseness, we suppressed factors such as $n_s$, $S_0$, etc. 
			
			A convenient way to carry out the analysis is to employ a coordinate change that transforms a modulated boundary into a straight one. This can be done 
			by a conformal mapping defined by an analytic function in the halfplane $y>0$ and an anti-analytic function in the halfplane $y<0$ with values matching at $y=0$. The most general function of this type is of the form 
			\be\label{eq:g_k}
			z'=x'+iy'=z+\sum_{k} g_k e^{ikx-|k| |y|}.
			\ee
			Under such conformal mapping the Schroedinger operator in Eq.\eqref{eq:action_wiggly} preserves its form up to a change in coefficients, allowing to describe a wiggly edge as a straight edge with a perturbation in the metric localized in its vicinity. The simple transformation rule is a consequence of the conformal invariance of the 2D Laplacian and the chiral density terms in our Hamiltonian, 
	Eq.(3) 
			in the main text. Indeed, denoting the Jacobian of the mapping in Eq.\eqref{eq:g_k} as $D(r)=(\p x', \p y')/(\p x, \p y)$, we find that under the conformal mapping the terms $ i\p_t-h_0$ in Eq.\eqref{eq:action_wiggly} are multiplied by $D^{-1}$ whereas other terms remain unchanged.

			The lifetime can now be calculated from the selfenergy for the Greens function
			\[
			G=\frac1{\omega-H}=\frac1{\frac{\omega-h_0}{D(r)}+J(\p_x^2+\p_y^2)-im\delta(y)\p_x}
			\]
			Expanding the Jacobian in powers of the modulation amplitude $g_k$ gives  $D^{-1}(r)=1+\delta p(r)$ where $\delta p(r)=\sum_k kg_k e^{ik x-|k||y|} +{\rm c.c.} +O(g_k^2)$. 
			We can now rewrite the Greens function in terms of the Hamiltonian for the straight edge, $H_0=h_0-J(\p_x^2+\p_y^2)+im\delta(y)\p_x$, and the perturbation $\delta p(r)$ localized near the edge:
			\be\label{eq:metric}
			G=\frac1{\omega-H_0+(\omega-h_0)\delta p(r)}
			.
			\ee
			This expression is exact and can therefore be used to obtain the lifetime of the chiral edge mode in a closed form. Starting with a normalized wavefunction for the chiral mode derived above, $\left. |\psi^0_+\ra=e^{iqx} e^{-\lambda_q|y|}\lambda_q^{1/2}$, and calculating the lifetime from the selfenergy of $G$ found at second order in $\delta p(r)$ we find the decay rate
			\[
			\gamma=2\pi\sum_{\vec Q} {\rm Im} G_0(\omega,\vec Q) |\la e^{i\vec Q\vec r}|(\omega-h_0)\delta p(r)|\psi^0_+\ra |^2
			\]
			where $G_0=\frac1{\omega-H_0+i0}$ and $\vec Q$ is the bulk magnon momentum. For a simple order of magnitude estimate it will be sufficient to approximate the spectral function ${\rm Im} G_0(\omega,\vec Q)$ as that of the bulk magnon continuum, ${\rm Im} G_0(\omega,\vec Q)=\pi \delta(\omega-J\vec Q^2)$. Estimating this expression we find the decay rate that scales as
			\be \label{eq:gamma}
			\gamma\sim \lambda_q(\omega-h_0)^2
			\ee
			At small $q$ these quantities scale as $\lambda_q\sim |q|$, $\omega-h_0\sim q^2$ yielding the decay rate that vanishes in the long-wavelength limit as $\gamma\sim q^5$. The long lifetime arises as a combination of two effects. First, because of the Larmor theorem, in the small $q$ limit the mode frequency for both bulk and edge is pinned to $h_0$ regardless of the presence of an edge disorder. Second, because at small $q$ the edge mode has a large penetration length into the bulk, $1/\lambda$. As a result, the mode weakly overlaps with the edge roughness, which suppresses the edge-to-bulk scattering. 
			
\end{document}